\documentclass[aps,apl,twocolumn,groupedaddress]{revtex4}
\usepackage{graphicx}
\usepackage{array}
\newcolumntype{P}[1]{>{\centering\arraybackslash}p{#1}}
\input epsf
\usepackage{amsmath,amssymb}

\begin{document}

%\title{STM/S study of variation in charge-state of interface defects in graphene FETs}
\title{STM/S study of electronic inhomogeneity evolution with gate voltage in graphene: role of screening and charge-state of interface defects}
\author{Anil Kumar Singh}
\affiliation{Department of Physics, Indian Institute of Technology Kanpur, Kanpur 208016, India}
\author{Anjan K. Gupta}
\affiliation{Department of Physics, Indian Institute of Technology Kanpur, Kanpur 208016, India}
\date{\today}

\begin{abstract}
Evolution of electronic inhomogeneities with back-gate voltage in graphene on SiO$_2$ was studied using room temperature scanning tunneling microscopy and spectroscopy. The reversal of local contrast in some places in the STS maps and sharp changes in cross-correlations between topographic and conductance maps, when graphene Fermi energy approaches its Dirac point, are attributed to change in charge-state of interface defects. The spatial correlations in the conductance maps, described by two different length scales and their growth during approach to Dirac point, show a qualitative agreement with the predictions of the screening theory of graphene. Thus a sharp change in the two length-scales close to the Dirac point, seen in our experiments, is interpreted in terms of the change in charge state of some of the interface defects. A systematic understanding and control of the charge state of defects will help in memory applications of graphene.
\end{abstract}
\maketitle

\section{Introduction}
Potential landscape created by charged defects buried in the substrate and at graphene-substrate interface leads to inhomogeneities in the carrier density in graphene devices. Such inhomogeneities have been observed experimentally in graphene on SiO$_2$ substrate by scanning single-electron transistor or by scanning tunneling microscopy and spectroscopy (STM/S) \cite{Martin-2008,Zhang-2009,Deshpande-2009,Deshpande-2011}. Such inhomogeneity is undesired as it limits the carrier mobility and restricts one from precisely accessing the Dirac point due to electron-hole puddle formation \cite{Mayorov-2012}. As a result several research groups have moved away from devices having graphene-SiO$_2$ interface as amorphous SiO$_2$ shelters many defects, particularly near the surface where graphene is placed. As a result of change in charge state of these defects, and/or due to the trapping of extrinsic species at the graphene-SiO$_2$ interface, the field electron devices with SiO$_2$ gate dielectric show significant hysteresis. The hysteresis, on the other hand, offers application potential in data storage. Thus understanding these interface charge traps, particularly for finding ways to control and probe the charge stored in them, is important.

Zhang et. al. \cite{Zhang-2009} found that the presence of charge-donating impurities below graphene caused fluctuations in graphene carrier  density due to formation of standing wave patterns by backscattering of Dirac fermions. Gibertini et. al. \cite{Gibertini-2012} reported that corrugations in graphene are sufficient for the formation of electron-hole puddles. Dielectric screening properties of graphene have been discussed by several groups \cite{Ando-2006, Katsnelson-2006, Adam-2007, Hwang-2009} to understand electron-hole puddle formation. Self consistent linear screening theory with random phase approximation (RPA) \cite{Adam-2007,Adam-2011} has been used \cite{Samaddar-2016} to model the observed growth, by STS, in the length-scale of charge inhomogeneities in graphene. Finite temperature also affects screening properties due to thermally activated electron-hole pair creation. Thus at room temperature the screening properties for carrier density below $10^{11}$ cm$^{-2}$ will differ from those of zero temperature as found from the linear RPA and non-linear Thomas-Fermi models \cite{Ghaznavi-2010}. The defects can also change their charge-state by exchanging electrons with graphene and can thus affect the carrier density as well as inhomogeneities. Depending on the overlap of defect states with graphene and temperature, this electron transfer can be dominated either by thermal activation or tunneling.

In this paper we report on the role of back-gate dependent charge-state of interface defects, other than the already known screening properties of graphene, in the evolution of electronic inhomogeneities in SLG on SiO$_2$ with back-gate-voltage ($V_g$). The evolution of the conductance and topographic maps of SLG on SiO$_2$ with $V_g$ is studied using a room temperature vacuum STM. The cross-correlations between topography and conductance maps of different $V_g$ show abrupt changes as $E_F$ approaches Dirac point in addition to a visible increase in the length-scale associated with charge inhomogeneity, in agreement with screening theory of graphene. From the evolution of local correlations in conductance maps with gate-voltage we clearly see some of the regions reversing their contrast and some preserving it. Further, the conductance maps are described by two length-scales and both grow as $E_F$ approaches Dirac Point with some abrupt changes. The later is discussed in terms of the change in the charge state of some of the defects.

\section{Experimental Details}

Graphene flakes were mechanically exfoliated from kish graphite using an adhesive tape on 300 nm thick SiO$_2$ on top of a highly n-doped silicon substrates. Optical contrast and Raman spectra of graphene flakes on SiO$_2$/Si substrate were used to identify the single layer graphene. A mechanical wire masking method was used to make contacts of Cr(10nm)/Au(50nm) on graphene. We used a homemade room temperature STM, with a 2D nano positioner \cite{Gupta-2008} sample holder, kept inside a chamber pumped by a cryo-pump with pressure maintained in $10^{-4}$ mbar range throughout the measurement. The gate-voltage was applied on Si substrate with a 10k$\Omega$ series resistance. Details of the device fabrication and measurement are same as those described elsewhere \cite{Singh-2016}.

We performed STM/STS studies on two different monolayer graphene samples (1 and 2) in a region away from the metal-graphene contact interface, both showing qualitatively similar behavior. Small-scale, see Fig. \ref{fig:spectra}(a), topographic image of this sample shows atomically resolved surface. The $V_g$ dependent local tunnel-conductance spectra at a point in fig. \ref{fig:spectra}(c) show two minima due to tip-doping effects \cite{Samaddar-2016,Singh-2016,Choudhary-2011,Zhao-2015}. The two move towards each other when gate voltage increases from -20V to 50V; they eventually merge close to V$_g=50$V implying significant p-doping with hole density of order 4$\times$10$^{12}$ cm$^{-2}$. Qualitatively similar spectra were seen throughout the sample with the major difference being the $V_b$-position of the minima. Similar doping is seen in the two probe resistance of an identically prepared sample in fig.\ref{fig:spectra}(b).

\subsection{Interpretation of STS maps}
Potential or carrier density inhomogeneity can be mapped in several ways using STS. One way is to acquire full tunneling conductance spectrum at each pixel with open-feedback in current imaging tunneling spectroscopy mode, which is comprehensive but demanding in terms of amount of data and time taken. We acquired spatial maps of dI/dV at a fixed bias (V$_b$) and in a closed feedback-loop, with large integration-time, using Lock-in method with an ac-modulation (2.731 kHz, 20mV rms). In this method, if we assume dI/dV-$V_b$ spectrum of graphene in the vicinity of V$_b$ to be monotonically rising, a small spatial variation in the position of the DP energy will lead to a shift in the primary minimum along the $V_b$ axis leading to a change in measured dI/dV value as elaborated further.

\begin{figure}
\includegraphics[width=8.3cm]{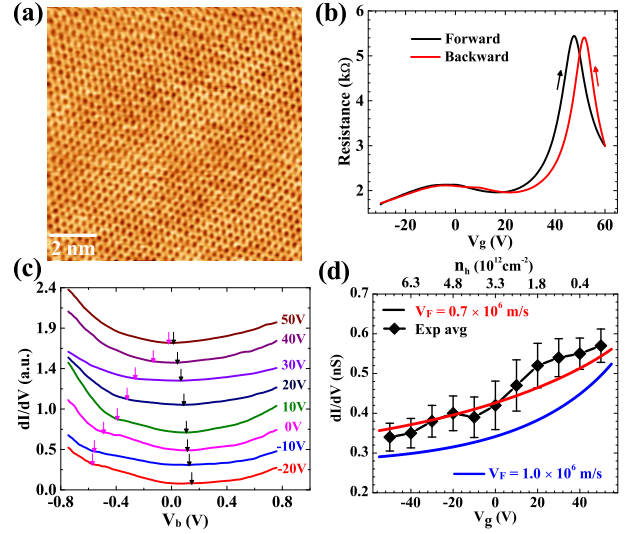}
\caption{(a) An STM image (11.2$\times$11.2 nm$^2$, 0.3 V/0.1 nA) of sample-1 showing atomically resolved honeycomb lattice of graphene. (b) shows the two probe resistance variation with $V_g$ on a graphene sample prepared using identical process. (c) shows the local tunneling conductance spectra (0.8V/0.1nA) at different $V_{g}$ values taken at a fixed location on graphene with black (purple) arrows marking the primary (secondary) minima. In (d) the continuous lines show the calculated dependence of conductance on $V_g$ (see text for details) while the discrete points (diamonds) show the average conductance found from STS maps with the bars depicting the standard deviation of the maps.}
\label{fig:spectra}
\end{figure}
From the spectra in fig. \ref{fig:spectra}(c), the primary minima movement in this sample occurs between 0 and 0.2 V and secondary minima remains at negative $V_b$ for the studied $V_g$ range. So, for STS maps, we chose $V_b=0.25$ V to avoid either of the two minima crossing this bias value for studied $V_g$ values. We approximate the low bias portion of the differential tunnel conductance by a parabola, i.e. $G(V_b,V_D)=G_0[1+G_1(V_b-V_D)^2]$. Here $V_D$ is the location of the primary minima with $eV_D$ as the energy of the Dirac point from the Fermi energy. Thus $V_D>0$ represents a hole doped graphene spectrum. $G_0$, $G_1$ are found by fitting the bottom portion of the local spectra \cite{suppl-info}. The conductance maps are acquired in closed-feedback-loop mode with fixed tunnel current and $V_b$ values. With feedback on, the tunnel current is kept constant (0.1nA) and so the variation in local conductance is assumed to occur due to $V_D$ variation. Thus we can write the measured local conductance in STS mode as $G_{STS}(V_D)=\frac{I(V_b,0)}{I(V_b,V_D)}G(V_b,V_D)$ with $I(V_b,V_D)=\int_0^{V_b}G(V,V_D)dV$. Further, attributing the $V_D$ variation to $V_g$, $V_b$ and a local effective potential, $V_{g}^{D}$, due to charged-defects and screening, we get
\begin{equation}V_D=\hbar v_F\sqrt{\frac{\pi\kappa\epsilon_0}{ed}}\sqrt{\frac{d}{\kappa z}V_b-(V_g-V_{g}^{D})}\label{eq:vd-vg}\end{equation} with $\kappa=4$ and $d=300nm$ as the dielectric constant and thickness of SiO$_2$ layer, respectively. We use $V_{g}^{D}$=50V, close to the doping seen in spectra and in resistance in fig.\ref{fig:spectra}(b). The tip-sample separation $z$ (in eq.\ref{eq:vd-vg}) is deduced to be 0.92 nm from the $V_g$ dependence of the secondary minima in fig. \ref{fig:spectra}(c). Fig.\ref{fig:spectra}(d) shows the conductance, thus calculated, for $v_F=1\times10^6$ and $0.7\times10^6$ m/s, together with the discrete $V_g$ dependent conductance and its spread, found from the spatial average of the conductance maps (presented later) for comparison. An apparent reduction in $v_F$ can be attributed to screening of gate electric field by the interface defects \cite{Singh-2016}.

With a fixed $V_{g}^{D}$ value, the above calculated conductance variation with $V_g$ ignores any $V_D$ change due to the change in charge-state of the defects. The electronic inhomogeneity, attributed to $V_{g}^{D}$ variation, gives rise to a spread in conductance as depicted in fig. \ref{fig:spectra}(d). The continuous line in this plot can be used to convert the conductance maps to carrier density maps, see the top scale of this plot; however, since this plot is monotonically rising, the conclusions drawn using the spatial correlations and associated length scales of conductance will not be affected. On the other hand, this plot gives a method to interpret conductance contrast, such as the nature, i.e. electron- or hole-type, of charge puddles and their evolution with $V_g$.

\section{STS study of electronic inhomogeneity}
\begin{figure}
\includegraphics[width=8.0cm]{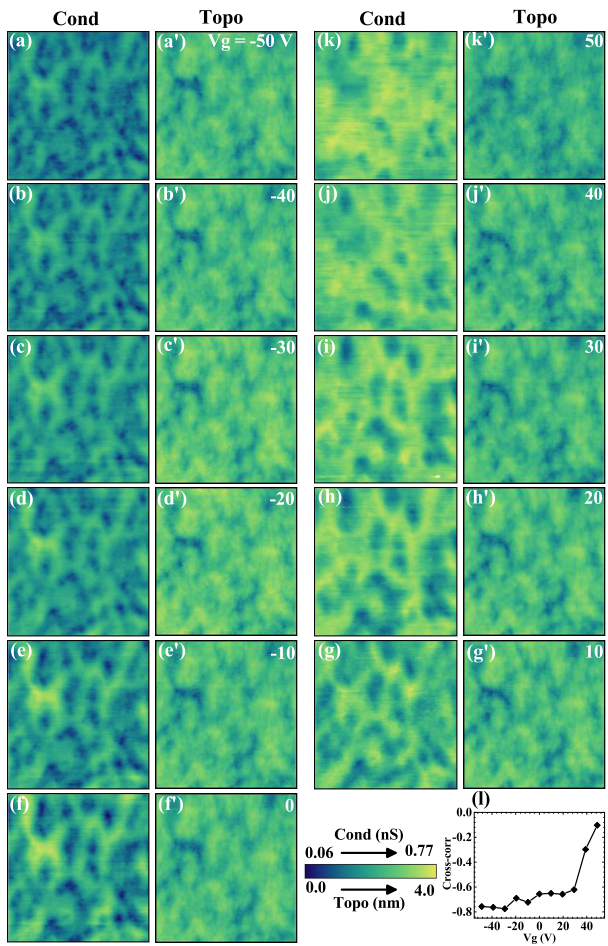}
\caption{(a) to (k) show the conductance maps depicting electron inhomogeneities with (a') to (k') showing corresponding topographic maps over an area of 120 $\times$ 120 nm$^2$ at different $V_g$ values. Imaging parameter for all the maps are bias voltage $V_{b}$ = 0.25V and current set-point 0.1 nA. (l) shows the cross-correlation coefficient between the conductance and conductance maps as a function of $V_g$.}
\label{fig:maps}
\end{figure}
Figure \ref{fig:maps} shows the STS conductance maps at various gate voltages between $\pm50$V in fig. \ref{fig:maps} together with the simultaneously taken topography images. The small scale topographic images in the same region, see fig. \ref{fig:spectra}(a), show the honeycomb structure of graphene. All the images in fig. \ref{fig:maps} correspond to the same area of the sample. This was ensured by finding the small relative shifts between topographic images of different $V_g$ by using the location of the peak in cross-correlation maps. The largest common area was then cropped from both the topography and conductance maps taken at different $V_g$. It can be seen from fig. \ref{fig:maps} that the topography images do not change noticeably with $V_g$ while there are significant changes in the conductance images, particularly when $V_g$ approaches +50 V, see fig. \ref{fig:maps}(a) and (k). This implies that the contribution to topographic contrast from electronic inhomogeneity is insignificant and these images reflect the actual topography of the surface as arising from the underlying SiO$_2$ \cite{Geringer-2009,Cullen-2010}. On the other hand, there is clear (anti-)correlation between topography and conductance images, see fig.\ref{fig:maps}(l), particularly for negative $V_g$ values, implying that the underlying topography is responsible for some of the electronic contrast \cite{Gibertini-2012}.

As per the plot in fig.\ref{fig:spectra}(d) the dark regions of an STS image represent high hole-density regions. Thus for the STS maps close to $V_g=50V$, when graphene $E_F$ coincides with Dirac point on average, the dark (bright) regions would represent hole (electron) puddles. Also at large positive $V_g$ the interface defects will have a tendency to acquire negative charge. Thus the positively charged defects will tend to become neutral (or negative) and the neutral ones, negative. The graphene region close to such defects will have carriers of opposite charge as compared to average charge density. Thus the electron puddles (above the +ve defects) will have a tendency to disappear (or change to hole puddles) as $V_g$ approaches +50V and some of the average carrier density regions (above neutral defects) will have a tendency to become hole puddles. The hole puddles (above -ve defects) are unlikely to change in this sense. One can similarly construct a reverse argument when $V_g$ goes to -50V. It is likely that a good number of defects have their ionization energies pinned close to the Dirac point of graphene \cite{Singh-2016,Romero-2008,Miwa-2011,Terrs-2016} and thus the change in their ionic state will occur when $V_g$ is close to 50V.
\begin{figure}
\includegraphics[width=8.3cm]{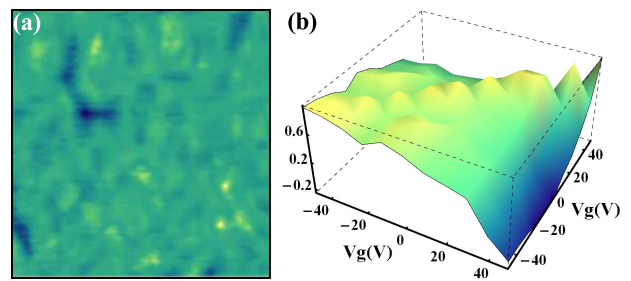}
\caption{(a) shows the product image of the conductance maps (after average subtraction and division by standard devication of respective maps) of $V_g=+50$ and -50 V. (b) shows a 3D rendering of cross-correlation coefficients between different $V_g$ conductance maps as a function of gate voltage.}
\label{fig:histogram}
\end{figure}

Sample-2 shows similar behavior in terms of local contrast evolution in STS maps with $V_g$ \cite{suppl-info}. In addition, in this sample we find that the STS maps show some sort of instability in a broad vicinity of the point when $E_F$ coincides with Dirac point. In fact the quality of STS maps in sample-1 close to $V_g=50$ V is also not as good. On the other hand the topographic image quality stays quite the same. This can occur due to tip-induced change in the charge-state of interface defects. The tip, with bias $V_b$, can play a significant role in this charge-state change by way of tip-doping effects. The stability of the STS images will be dictated by the competition between the rate of charge-state change and the tip scan rate. The former rate can, in fact, be very slow \cite{Xu-2012} and thus slowing down the scan speed may not be a practical solution for getting stable images.

We can see several regions in the STS images in fig. \ref{fig:maps} that retain their contrast even up to $V_g=+50$ V while there are some that change to opposite contrast. This is better seen in fig. \ref{fig:histogram}(a) which shows the product of the +50 and -50 V maps after average subtraction and division by standard deviation. The average of this product map gives the cross-correlation of the two maps [$A(0,0)$ with respect to eq.\ref{eq:cross}], which is close to -0.2, see fig. \ref{fig:histogram}(b). From the same plot the cross correlation between different STS maps is seen to be more than +0.5 except for $V_g=40$ and 50 V. This abrupt jump around $V_g=40$ V is again believed to occur due to defect-state change. There are some prominent bright and dark regions in this product map with the former (later) indicating position of defects that maintain (change) their charge state. The dark (anti-correlated) regions of the product map can arise either from bright becoming dark or from dark becoming bright; as seen from fig. \ref{fig:maps}(a) and \ref{fig:histogram}(a), bright becoming dark dominates. Also the bright regions of the product map can come from bright remaining bright or dark remaining dark and it is the later which occurs almost exclusively.

We interpret this as the hole puddles (dark regions) that exist above negatively-charged defects retaining their character while the electron puddles (bright regions) that exist above positively charged defects change their character when gate voltage approaches +50V. This is consistent with earlier discussed expectation on change in charge-state of defect with gate voltage. There are some dark regions (hole puddles) that also disappear due to some negative defects getting neutralized. This is a very qualitative approach and more detailed evolution of this contrast will depend on the detailed distribution of the two types of defects. The evolution of contrast and the size of the electron and hole puddles will also depend on the screening properties of graphene which we discuss later. The sample-2 \cite{suppl-info} shows similar local contrast evolution in STS maps.

We would also like to point out that for sample-1 the STM/S images were taken in the $V_g$ sequence +40 to -50 V and then +50 V, i.e. fig. \ref{fig:maps}(j) to (a) and then (k). Looking at the continuity between the +40V and +50V STS images taken after full $V_g$ cycle we do not see any visible signatures of hysteresis, which is anyway seen to be quite small from the transport measurements, see fig. \ref{fig:spectra}(b), on an identically prepared sample. The hysteresis is believed to arise from the meta-stability of the charge state of defects near graphene-SiO$_2$ interface \cite{Xu-2012}. Our STS measurements are quite slow in the sense that each image takes about an hour to complete. In addition, the tip (with a bias voltage) can also help in relaxing the local defects out of the meta-stable state.

\begin{figure}
\includegraphics[width=8.0cm]{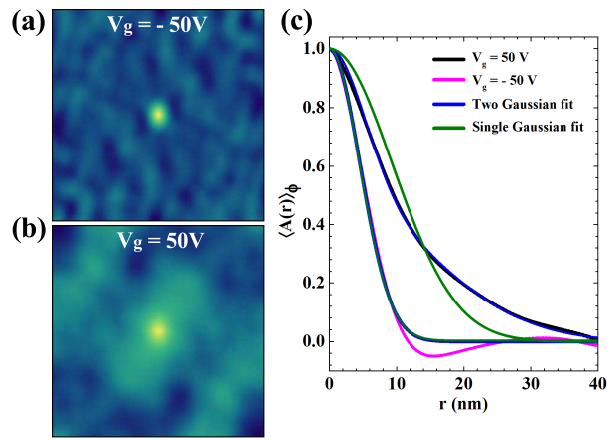}
\caption{(a) and (b) show the autocorrelation image calculated using eq.\ref{eq:cross} of conductance map at $V_g=-50$ and +50 V, respectively. (c) shows angular average of the line cut passing through the center of (a) and (b) with its single and double Gaussian fits.}
\label{fig:analysis}
\end{figure}

\subsection{Puddle size evolution}
To quantify the puddle size and analyze its evolution with V$_g$ we filtered all the images using a Gaussian filter of 1 nm width before calculating correlations. We find the cross correlation $A(x,y)$ between two experimentally acquired images, i.e. $z_{1,2}(i,j)$, using,
\begin{equation}
\small A(i,j)=\frac{\sum\limits_{i',j'}^{} z_1 (i', j') z_2 ( i+i', j+j') - \langle z_1\rangle \langle z_2\rangle}{\sqrt{\left[(\langle z_1^{2}\rangle - \langle z_1\rangle^2)(\langle z_2^{2}\rangle - \langle z_2\rangle^2 )\right]}},
\label{eq:cross}
\end{equation}
Here the sum over $i'$ and $j'$ and the averages ($<..>$) are evaluated over the overlapping area after the relative shift by $i$ and $j$. The same is used for finding the auto-correlation. With this expression the cross-correlation will be normalized, i.e. it will give a value +1 for perfect correlation, zero for no correlation and -1 for perfect anti-correlation. Thus an auto-correlation image will have value +1 at the origin. A Mathematica program was written to calculate the auto-correlation images using eq.\ref{eq:cross}.

Line-cut in this cross correlation function A$(x,y)$ through the origin can be represented by \emph{A(r,$\phi$)} along a fixed direction $\phi$. Angular averaged, normalized correlation function $\langle A(r) \rangle_{\phi}$ can be fitted to a Gaussian function or a combination of Gaussian functions of the form $exp[-r^2/2\xi^2]$ with $\xi$ as correlation length. Fig.\ref{fig:analysis} (a) shows the auto-correlation image of dI/dV image at V$_g$ = -50 V. From this image it is clear that it has a sharp peak at $r=0$ with certain decay length $\xi$. The $V_g=50$V autocorrelation in fig.\ref{fig:analysis} (b) shows a sharp decay, i.e. a small correlation length, followed by a gradual decay, i.e. a large correlation length. This indicates two length scales. This is clear from $\langle A(r)\rangle_{\phi}$ shown in Fig.\ref{fig:analysis} (c). We have fitted this $\langle A(r)\rangle_{\phi}$ with single Gaussian function $exp[-r^2/2\xi^2]$ as well as with a sum of two Gaussian functions, i.e., \begin{equation}\sigma exp[-r^2/2\xi_{1}^{2}] + [(1-\sigma)exp[-r^2/2\xi_{2}^{2}]\label{eq:2-gaussian}.\end{equation} At V$_g=-50$ V, the $\langle A(r)\rangle_{\phi}$ fits well with a single Gaussian whereas for V$_g=50$ V it has large deviation from a single Gaussian. But both the curves can be well fitted by sum of two Gaussian functions.

The presence of two length scales near Dirac point can be seen directly from the conductance map in fig. \ref{fig:maps}(k) from a few scattered small-size dark-spots in a relatively flat and bright looking background. The dark spots can arise from segregated defect clusters giving small hole puddles in a well connected electron-like landscape. Such two different size puddles in potential landscape are also anticipated in an effective medium theory \cite{Rossi-2009}. The presence of two length scales near Dirac point is also found to be valid for the STS images of sample-2 \cite{suppl-info}, as per fig.\ref{fig:analysis}(b), in which the charge neutrality point appears close to $V_g=20$ V. In order to discuss the $V_g$-evolution of these length scales we first briefly discuss the predictions of linear screening theory of defect potential in graphene.

\section{linear screening theory of electronic inhomogeneity}

In graphene, the fine structure constant, (r$_s$ = 0.8 for graphene deposited on SiO$_2$) is independent of carrier concentration, implying that the strength of electron-electron interaction remains fixed and it is a weakly interacting system due to its linear dispersion. The Thomas-Fermi wave vector q$_{TF}$ is proportional to the square root of the carrier density. As a consequence unscreened potential created by charge impurities and screened potential are identical near DP and screening length q$_{TF}^{-1}$ strongly depends on the carrier density.

For a random distribution of charged impurities with density $n_{imp}$ in a plane at distance $d$ from the graphene sheet, the screened impurity potential is given by \cite{Adam-2007},
\begin{equation}
\small C(r)=2\pi n_{imp}\left(\frac{e^{2}}{4\pi\epsilon_{0}\kappa}\right)^{2}\int_{0}^{\infty}
\left[\frac{1}{\epsilon(q)}\frac{e^{-qd}}{q}\right]^{2}J_{0}(qr)qdq.
\label{eq:SDP}
\end{equation}
Here $\kappa$ is the bulk (3D) dielectric constant, $J_{0}$ is the zeroth-order Bessel function, $e$ is the magnitude of electronic charge and $\epsilon(q)$ is the temperature dependent graphene dielectric function. Normalized correlation function $A(r)=C(r)/C(0)$ is more useful for describing the spatial profile of the screened impurity potential or electron-hole puddles while $C(0)$ characterizes the average potential fluctuations i.e. \~{V}$_{rms}$. $A(r)$ has a Gaussian-like appearance, i.e. $exp[-r^2/2\xi^2]$ with $\xi$ as a correlation length.

\begin{figure}
\includegraphics[width=8.7cm]{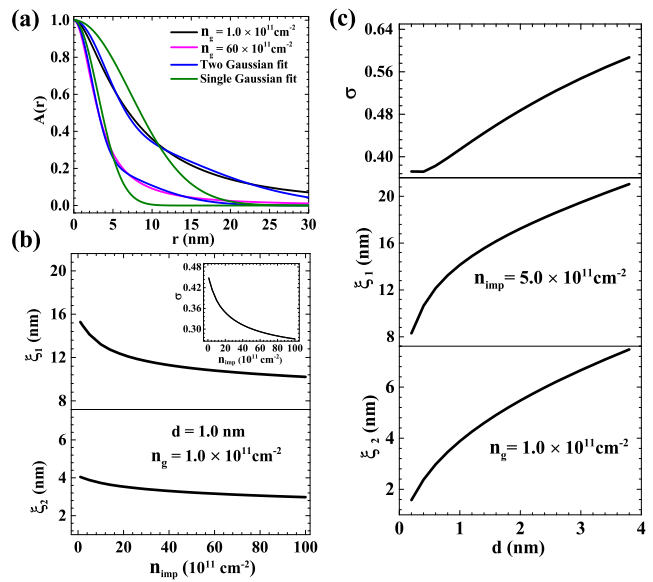}
\caption{(a) Single and two Gaussian fits of theoretically calculated A(r) at $d=1.0$ nm and $n_{imp}=5\times 10^{11}$ cm$^{-2}$ for low $n_g$ ($\xi=7.2$ nm, $\sigma=0.41$, $\xi_1=14.1$ nm, $\xi_2=3.9$ nm) and high $n_g$ ($\xi=3.1$ nm, $\sigma=0.24$, $\xi_1=7.9$ nm, $\xi_2=2.2$ nm).(b) shows the evolution of $\xi_1$ and $\xi_2$ as a function of $n_{imp}$ at $d=1.0$ nm; inset shows $\sigma$ as a function of $n_imp$. (c) shows the evolution of $\sigma$, $\xi_1$ and $\xi_2$ as a function $d$ at $n_{imp} = 5\times10^{11}$ cm$^{-2}$.}
\label{fig:model-plots}
\end{figure}

We calculated $A(r)$ using the $\epsilon(q)$ corresponding to linear screening theory with random phase approximation \cite{Adam-2007,Hwang-2009,suppl-info} for fixed $T=300$ K to find how the correlation length(s) vary with $n_{imp}$, $d$ and $n_g$. Fig. \ref{fig:model-plots}(a) shows $A(r)$ for $n_{imp}=5\times10^{11}$ cm$^{-2}$, $d=1.0$ nm and for two different carrier density ($n_g$) values together with their fits to single and double Gaussian (see eq. \ref{eq:2-gaussian}). Clearly the double Gaussian fit is much better, which is found to be the case for a wide range of $n_{imp}$, $d$ and $n_g$ values. The discrepancy of $A(r)$ with its single Gaussian fit was found to be more pronounced at small $n_g$ and large $d$ values. In fact, the effective medium theory, near Dirac point, by Rossi et. al. \cite{Rossi-2009}, which includes the nonlinear screening and exchange-correlation effects, also finds two distinct length scales in the screened potential.

Fig. \ref{fig:model-plots}(c) shows the calculated variation of $\xi_{1}$ and $\xi_{2}$ with n$_{imp}$ at small $n_g$. Both lengths decrease with increase in $n_{imp}$ but the variation in $\xi_{2}$ with $n_{imp}$ is much smaller as compared to that in $\xi_{1}$ for a given $d$. Fig. \ref{fig:model-plots}(d) shows $d$-dependence of $\xi_{1,2}$, at small $n_g$, with both showing a significant increase with $d$. $n_g$ dependence of $\xi_{1,2}$ \cite{suppl-info} for different $n_{imp}$ and $d$ shows a decrease in both as we move away from DP. Mathematically $A(r)$, and thus $\xi_{1,2}$, are functions of two independent variables, namely $d$ and $q_{TF}^{-1}$, with the later dependent on $n_{imp}$ and $n_g$. Not knowing the functional form of dependence of $\xi_{1,2}$ dependence on $d$ and $q_{TF}^{-1}$, we discuss it here qualitatively. $\xi_1$ has a noticeable dependence on all the three parameters, i.e. $d$, $n_g$ and $n_{imp}$, while $\xi_2$, on the other hand, has noticeable dependence on $d$ but it has relatively weak dependence on $n_g$ and $n_{imp}$. $n_g$ and $n_{imp}$ play somewhat similar role as both dictate the average carrier concentration \cite{Adam-2007}.

\begin{figure}
\includegraphics[width=8.5cm]{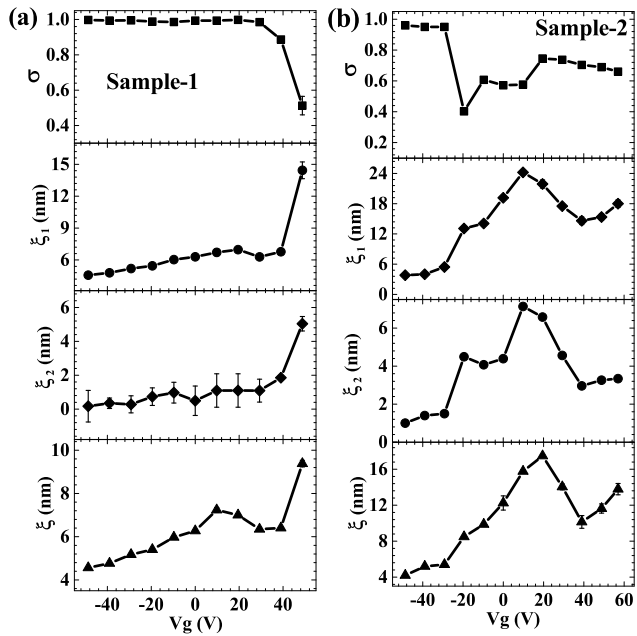}
\caption{Evolution of $\sigma$, $\xi_1$, $\xi_2$ and $\xi$ as a function of gate voltage $V_{g}$ extracted from conductance maps of (a) sample-1, and (b) sample-2.}
\label{fig:summary}
\end{figure}

\section{$V_g$ evolution of $\xi$: screening and defects' charge-state}

Fig.\ref{fig:summary} (a) shows the $V_g$ dependence of $\xi_{1,2}$, $\sigma$ and $\xi$ as found using the STS images, shown in fig. \ref{fig:maps}, of sample-1. Here $\xi$ corresponds to single Gaussian fit. $\xi_1$ and $\xi_2$ found by fitting are differentiated by their magnitudes and smooth variation of each with $V_g$. When $\sigma$ takes values close to 0 or 1, the single Gaussian was found to fit $A(r)$ quite well. This can be easily understood from eq. \ref{eq:2-gaussian}. Away from DP single Gaussian with length $\xi$ fits $A(r)$ and $\xi$ grows when DP is approached. Similar behavior can be seen for sample-2 (with DP at $V_g\sim20$ V) in the sense that single Gaussian fits away from DP and $\xi_{1,2}$ or $\xi$ grow as $E_F$ approaches Dirac Point. In fact the variation of $\xi_{1,2}$, particularly in the regime where two Gaussians fit better, is rather abrupt as compared to what is expected from the model \cite{suppl-info}. Incidentally, the correlation length from the topography images is found to be close to 7 nm \cite{suppl-info} for all $V_g$ values.

The growth of $\xi$'s as one approaches Dirac point, which is also clearly seen in the STS images, is in qualitative agreement with screening theory predictions. However, the abrupt changes in $\xi$ when $E_F$ approaches Dirac point are not quite anticipated. Some fluctuations in $\xi$ can be understood given that we are sampling relatively small area. However, these abrupt changes appear rather systematic and can arise from the change in charge-state of defects as this will change $n_{imp}$, $n_g$, as well as average $d$ as mainly the impurities close to graphene will change their state. Samaddar et al. \cite{Samaddar-2016}, from their STM/S study, have reported similar qualitative agreement with the screening theory together with some abrupt changes in $\xi$.

We have already discussed the possibility of interface defects changing their state from the evolution of various correlations between different STS or between STM and STS images. Since these samples are p-doped, due to interface defects, we expect majority of the interface defects to have negative charge. The presence of positive hysteresis \cite{Wang-2010} in fig. \ref{fig:spectra}(b) indicates that interface defects do change their charge state although converse need not be true. During forward sweep of $V_g$ the defects will have a tendency to change their charge state to more negative. One can also analyze this defect-state-change in terms of change in filling of the defect-states that are distributed with respect to the graphene Fermi energy\cite{Terrs-2016}. In equilibrium the defect states up to graphene Fermi energy will be filled \cite{Singh-2016}. Thus the interface states that can possibly change their filling due to $V_g$ change will be close to $E_F$ or the Dirac point, as the later is accessible in our samples by $V_g$ change. The overlap of interface states' wave function with graphene will dictate the actual equilibration time and the system may get trapped in a meta-stable state giving rise to hysteresis.

\section{Conclusions}

In conclusion, the interface defects, which are responsible for electronic inhomogeneities in graphene, are observed to change their charge state with change in back gate voltage and thus contribute to the evolution of graphene's electronic inhomogeneity other than the carrier density dependent screening physics. The change in defect state is predominantly seen when the graphene Fermi energy is close to the Dirac point implying that the energy-levels of the electrons bound to some of the defects are close to the Dirac point. Also, close to the Dirac point, where the screening of the impurity potential in graphene sheet is extremely weak, the electronic inhomogeneity in graphene is described by two different length scales and in our hole-doped samples we get small size electron puddles in a hole background varying over a much larger length-scale.

\section{Acknowledgement}

Financial support from the DST of the Government of India and from IIT Kanpur are gratefully acknowledged.


\begin{thebibliography}:
\bibitem{Martin-2008} J. Martin, N. Akerman, G. Ulbricht, T. Lohmann, J. H. Smet, K. von Klitzing and A. Yacoby, Nature Physics {\bf 4}, 144-148 (2008).
\bibitem{Zhang-2009} Y. Zhang, V. W. Brar, C. Girit, Alex Zettl and M. F. Crommie, Nature Physics {\bf 5}, 722-726 (2009).
\bibitem{Deshpande-2009} A. Deshpande, W. Bao, F. Miao, C. N. Lau and B. J. LeRoy, Phys. Rev. B {\bf 79}, 205411 (2009).
\bibitem{Deshpande-2011} A. Deshpande, W. Bao, F. Miao, C. N. Lau and B. J. LeRoy, Phys. Rev. B {\bf 83}, 155409(4) (2011).
\bibitem{Mayorov-2012}  A. S. Mayorov, D. C. Elias, I. S. Mukhin, S. V. Morozov, L. A. Ponomarenko, K. S. Novoselov, A. K. Geim and R. V. Gorbachev, Nano Lett. {\bf 12}, 4629 (2012).
\bibitem{Gibertini-2012} M. Gibertini, A. Tomadin, F. Guinea, M. I. Katsnelson and M. Polini, Phys. Rev. B {\bf 85}, 201405(R) (2012).
\bibitem{Ando-2006} T. Ando, J. Phys. Soc. Jpn. {\bf 75}, 074716 (2006).
\bibitem{Katsnelson-2006} M. I. Katsnelson, Phys. Rev. B {\bf 74}, 201401(R) (2006).
\bibitem{Adam-2007} Shaffique Adam, E. H. Hwang, V. M. Galitski, and S. D. Sarma, PNAS {\bf 104}, 18392–18397 (2007).
\bibitem{Hwang-2009} E. H. Hwang and S. D. Sarma, Phys. Rev. B {\bf 79}, 165404 (2009).
\bibitem{Adam-2011} S. Adam, S. Jung, N. N. Klimov, N. B. Zhitenev, J. A. Stroscio and M. D. Stiles, Phys. Rev. B {\bf 84}, 235421 (2011).
\bibitem{Samaddar-2016} S. Samaddar, I. Yudhistira, S. Adam, H. Courtois and C. B. Winkelmann, Phys. Rev. Lett. {\bf 116}, 126804 (2016).
\bibitem{Ghaznavi-2010} M. Ghaznavi, Z. L. Miškovic and F. O. Goodman, Phys. Rev. B {\bf 81}, 085416 (2010).
\bibitem{Gupta-2008} A. K. Gupta, R. S. Sinha, and R. K. Singh, Rev. Sci. Instrum. {\bf 79}, 063701 (2008).
\bibitem{Singh-2016} A. K. Singh and A. K. Gupta, arXiv:1612.08969v2(2016).
\bibitem{Choudhary-2011} S. K. Choudhary and A. K. Gupta, Appl. Phys. Lett.{\bf 98}, 102109 (2011).
\bibitem{Zhao-2015} Y Zhao, J. Wyrick, F. D. Natterer, J. F. Rodriguez-Nieva, C. Lewandowski, K. Watanabe,T. Taniguchi, L. S. Levitov, N. B. Zhitenev and J. A. Stroscio, Science {\bf 348}, 672-675 (2015).
\bibitem{suppl-info} Supporting Information
\bibitem{Geringer-2009} V. Geringer, M. Liebmann, T. Echtermeyer, S. Runte, M. Schmidt, R. Rückamp, M. C. Lemme and M. Morgenstern, Phys. Rev. Lett. {\bf 102}, 076102 (2009).
\bibitem{Cullen-2010} W. G. Cullen, M. Yamamoto, K. M. Burson, J. H. Chen, C. Jang, L. Li, M. S. Fuhrer and E. D. Williams, Phys. Rev. Lett. {\bf 105}, 215504
(2010).
\bibitem{Romero-2008} H. E. Romero, N. Shen, P. Joshi, H. R. Gutierrez, S. A. Tadigadapa, J. O. Sofo and P. C. Eklund, ACS Nano {\bf 2}, 2037 (2008).
\bibitem{Miwa-2011} R. H. Miwa, T. M. Schmidt, W. L. Scopel and A. Fazzio, Appl. Phys. Lett. {\bf 99}, 163108 (2011).
\bibitem{Terrs-2016} B. Terres, L. A. Chizhova, F. Libisch, J. Peiro, D. Jorger, S. Engels, A. Girschik, K. Watanabe, T. Taniguchi, S.V. Rotkin, J. Burgdorfer and C. Stampfer, Nat Commun. {\bf 7}, 11528 (2016).
\bibitem{Xu-2012} H. Xu, Y. Chen, J. Zhang, and H. Zhang, small {\bf 8}, 2833 (2012).
\bibitem{Rossi-2009} E. Rossi, S. Adam and S. D. Sarma, Phys. Rev. B {\bf 79}, 245423 (2009).
\bibitem{Wang-2010} H. Wang, Y. Wu, C. Cong, J. Shang, and T. Yu, ACS Nano {\bf 4}, 7221 (2010).

\end{thebibliography}
\end{document}

\title{Supporting Information on ``STM/S study of electronic inhomogeneity ......"}
\author{Anil Kumar Singh}
\affiliation{Department of Physics, Indian Institute of Technology Kanpur, Kanpur 208016, India}
\author{Anjan K. Gupta}
\affiliation{Department of Physics, Indian Institute of Technology Kanpur, Kanpur 208016, India}
\date{\today}

\maketitle
\section{Fitting of the spectra}
The fits of the low $V_b$ portion of some of the actual dI/dV spectra, taken from fig. 1(c), to a parabolic form $G(V_b)=G_0(1+G_1(V_b-V_D)^2)$ with $G_1=0.078$ V$^{-2}$ are shown in fig. \ref{fig:spec-fit} below. The $V_D$ value corresponds to the primary minimum and it depends on $V_g$. Since the actual conductance scale in the spectra depends on the current set point and bias, what matters here is $G_1$. The conductance scale in fig. 1(d) (of main text) is fixed by ensuring the current value at 0.25V bias as 0.1nA.
\begin{figure}
\includegraphics[width=8.0cm]{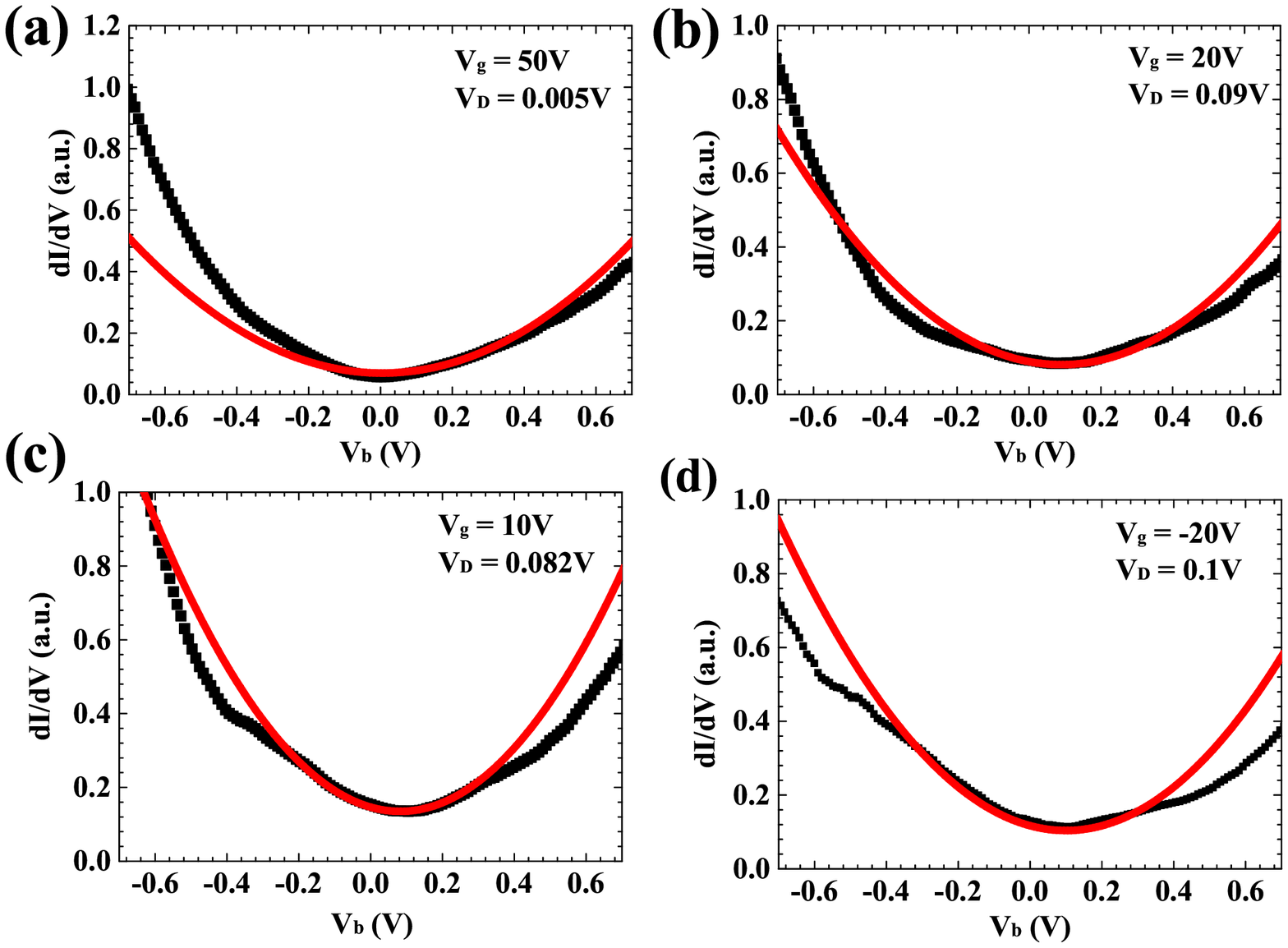}
\caption{Parabolic fits of the low bias portion of the local spectra (from fig. 1 of the main text) at several gate voltages.}
\label{fig:spec-fit}
\end{figure}

\section{STS maps of Sample-2}
\begin{figure}
\includegraphics[width=8.0cm]{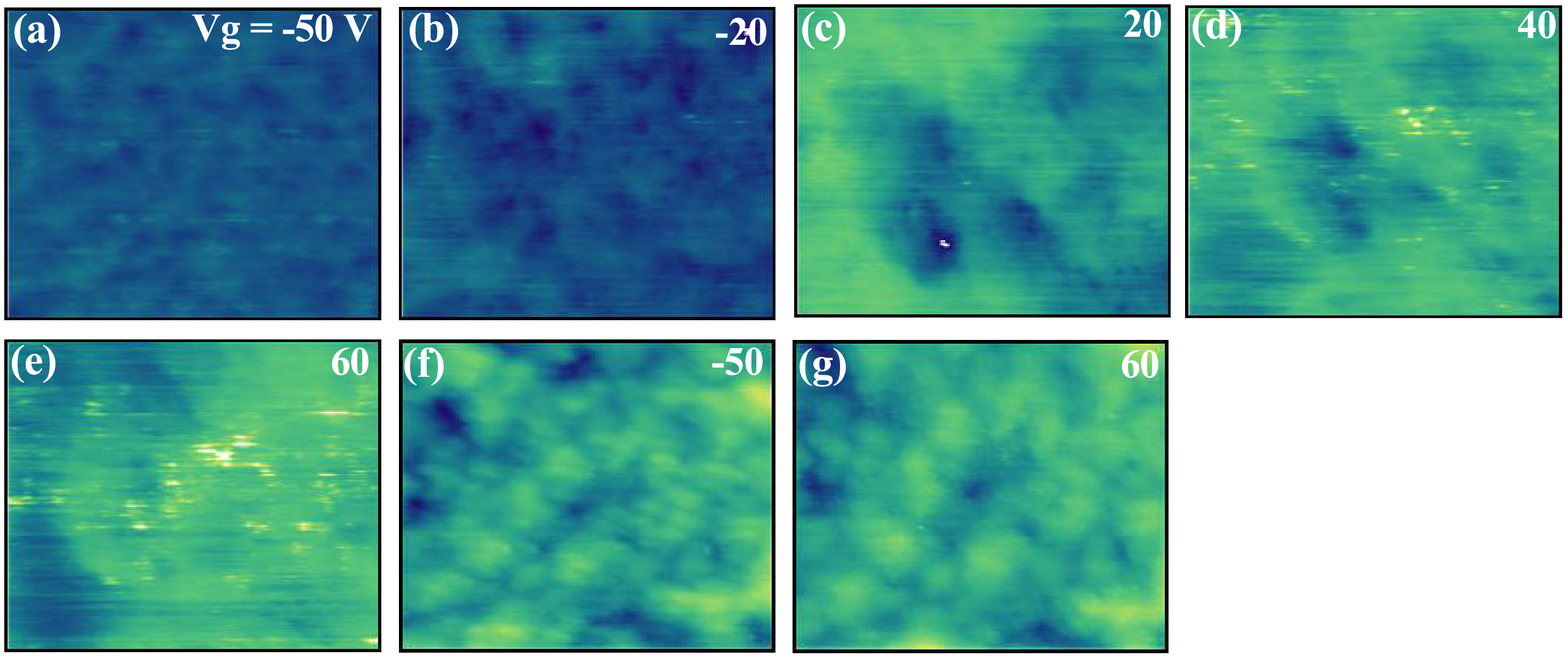}
\caption{ (a) to (e) show the spatial maps of electron hole puddles over an area of 135 $\times$ 110 nm$^2$ at different $V_{g}$ voltages. (f) and (g) showing corresponding topographic maps at two extreme $V_{g}$ (0.15 V/0.1 nA).}
\label{fig:maps2}
\end{figure}
All the conductance and corresponding topography images of sample-2 as shown in fig. \ref{fig:maps2} are from the same area. The evolution of these images with $v_g$ is quite similar to that of sample-1 in main text, such as 1) insignificant changes in topography and noticeable changes in the conductance images with $V_g$, 2) anti-correlation between topography and conductance images, and 3) evolution $\xi$'s as shown in fig. 6(b) of main text. Also there are three dark spots (near the center of conductance map) that persist beyond DP and up to $V_g=+50$ V.

\section{Dielectric function for linear self consistent screening theory}
The temperature dependent graphene dielectric function $\epsilon(q)$ is given by \cite{Hwang-2009},
\begin{equation}
\small \epsilon(q,T)= 1 + \frac{4 r_{s} k_{F}}{q}\tilde{\Pi}(q,T).
\label{eq:a}
\end{equation}
Here $r_{s}={e^{2}}/{4\pi\epsilon_{0}\kappa\hbar v_{F}}\approx 0.8$ on SiO$_2$ with $v_F$ as the Fermi velocity of graphene, $k_{F}=\sqrt{\pi n}$ with $ n= n_{g} + n^{*}$. Charge carrier density n$_g$ depends on $V_g$ and n$^{*}$ represents the disorder-induced residual carrier density when $n_g=0$ \cite{Adam-2007}. The dimensionless quantity ${\tilde{\Pi}(q,T)=\Pi(q,T)/N_0}$ with $\Pi(q,T)$ is the finite-temperature graphene polrizability with $N_0=2E_F/\pi(\hbar v_F)^2$ as the DOS at E$_F$. Within RPA we have\cite{Hwang-2009},
\begin{equation}
\begin{aligned}
\small \tilde{\Pi}(q,T) & =\frac{\mu}{E_{F}}+ \frac{\pi q}{8 k_{F}} + \frac{2T}{T_F}\ln(1 + e^{-\beta\mu})-\frac{1}{k_F}\times\\ & \left[\int_{0}^{q/2}\frac{\sqrt{1-(2k/q)^2}}{(1 + e^{\beta(\epsilon_k-\mu)})}dk + \int_{0}^{q/2}\frac{\sqrt{1-(2k/q)^2}}{(1 + e^{\beta(\epsilon_k +\mu)})}dk \right],
\end{aligned}
\label{eq:b}
\end{equation}
\begin{figure}
\includegraphics[width=8.0cm]{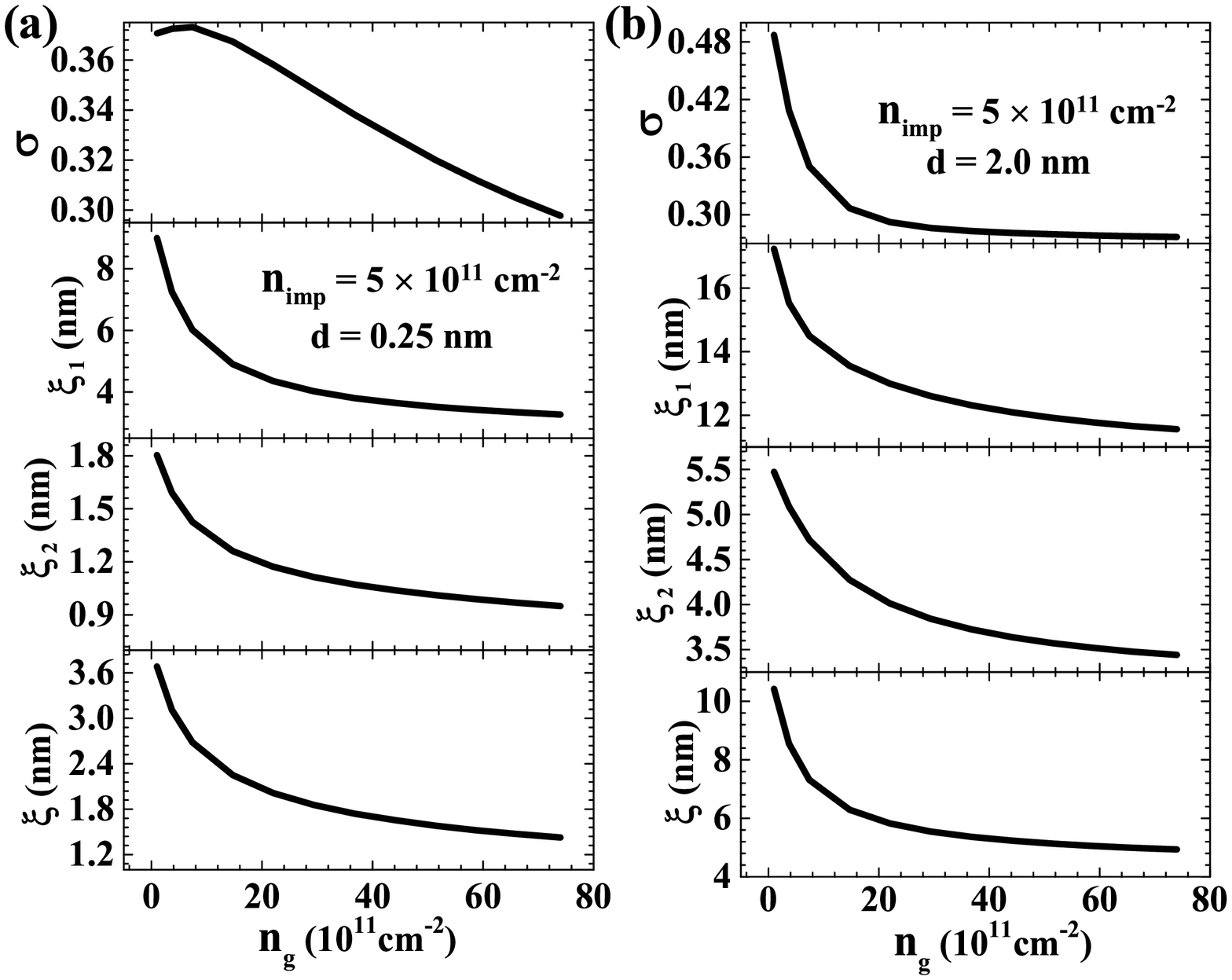}
\caption{ (a) shows the evolution of $\sigma$, $\xi_1$, $\xi_2$ and $\xi$ as a function of $n_g$ for d= 0.25nm and $n_{imp}=5\times10^{11} cm^{-2}$. (b) shows the evolution of $\sigma$, $\xi_1$, $\xi_2$ and $\xi$ as a function n$_g$ for d= 2.0nm and $n_{imp}=5\times10^{11} cm^{-2}$.}
\label{fig:theory}
\end{figure}
here $E_{F}=k_{B}T_{F}$ is the Fermi energy with respect to DP, $\epsilon_{k}=\hbar v_{F}k$, $\beta=1/k_{B}T$ and $\mu$ is the finite temperature chemical potential and it can be determined by the conservation of the total electron density. From screened impurity potential C(r) (see main text),
\begin{equation}
\small C(0)=2\pi n_{imp}\left(\frac{e^{2}}{4\pi\epsilon_{0}\kappa}\right)^{2}\int_{0}^{\infty}
\left[\frac{1}{\epsilon(q,n)}\frac{e^{-qd}}{q}\right]^{2}qdq,
\label{eq:ASDP}
\end{equation}
which characterizes the potential fluctuations i.e. \~{V}$_{rms}$. Quantitatively, n$^{*}$, defined as the charge carrier density at which the rms fluctuations in the screened disorder potential become equal to to the mean Dirac energy, can be calculated by self-consistent equation \cite{Adam-2007}, $\pi(\hbar v_F)^2 n^{*}= C(0)$.

Fig. \ref{fig:theory} shows the evolution of length scales as function of gate dependent carrier density $n_g$ for two different $d$ values. $\xi$'s decrease as we increase $n_g$. Here we want to point out that the actual puddle size observed in STS maps is larger than $\xi$ close to Dirac point. Since the single length scale fits gives the average puddle size but in actual puddle size have totally distinct two length scale. With d this discrepancy increases drastically because difference between two length scale become larger and can not be approximated as a single length scale in average sense also. Presence of isolated impurities at the dirac points which we have seen in both sample is responsible for two length scale as predicted by by Rossi et al.\cite{Rossi-2009}. In the absence of these isolated impurities people may find the puddle size much larger equal to sample size at the dirac point \cite{Martin-2008}.
\begin{figure}
\includegraphics[width=8.0cm]{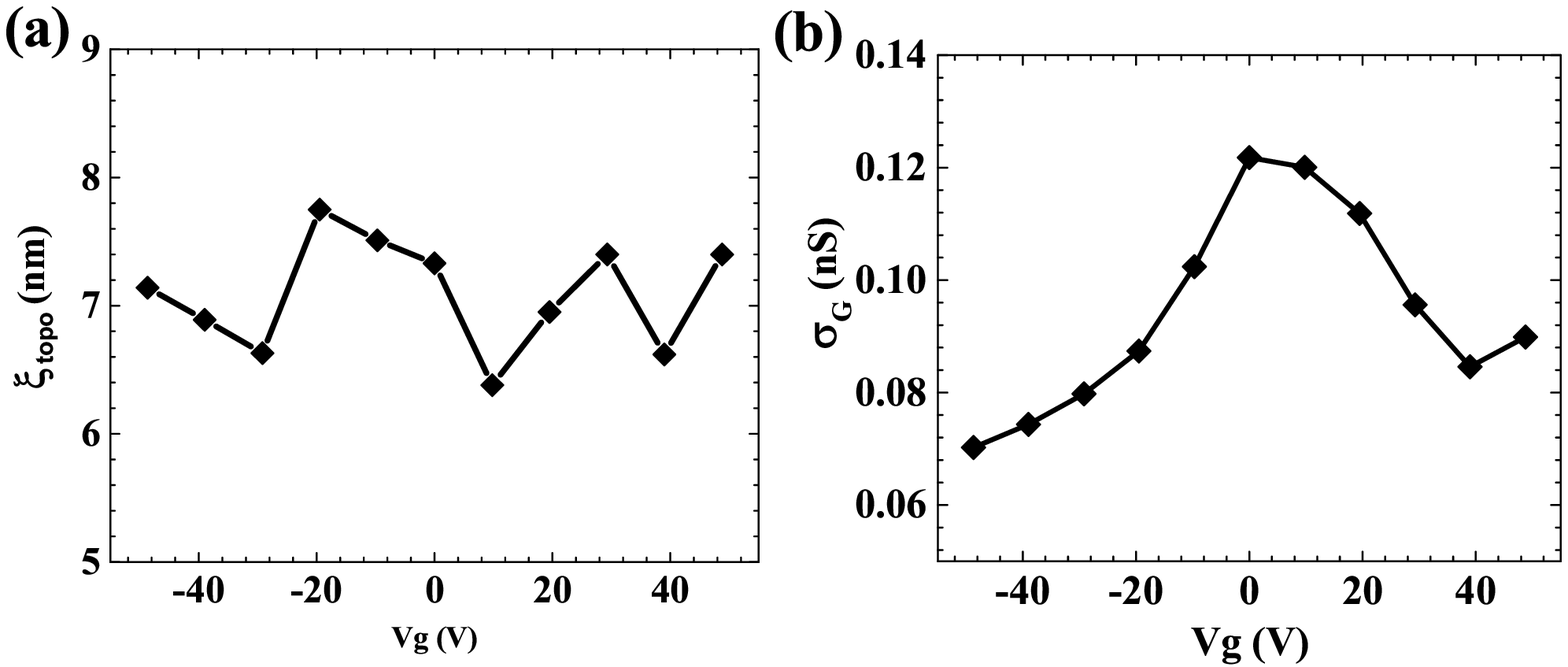}
\caption{ (a) Topography correlation length evolution with $V_g$ for sample-1. (b) shows the standard deviation of the STS maps of Sample-1.}
\label{fig:topo-avg}
\end{figure}

\section{$V_g$ evolution of topographic correlation length and rms conductance}

We found, in both samples, topography images do not change contrast and the correlation length, see fig. \ref{fig:topo-avg}(a) showing a fixed correlation length of $\simeq$ 7 nm, with $V_g$ while there are significant changes in the conductance images. Fig. \ref{fig:topo-avg} (b) shows the variation of rms conductance, equivalent to $C(0)$, with $V_g$ for sample-1. Thus $C(0)$ does not show a monotonic increase, as expected from screening theory, when $E_F$ approaches Dirac point and again, presumably, due to defect-state change.